\documentclass[letterpaper, 10pt, conference]{ieeeconf} 
\IEEEoverridecommandlockouts
\usepackage{amsmath}
\usepackage{amsfonts}
\usepackage{graphicx}
\usepackage{color}
\usepackage{subeqnarray}
\usepackage{algorithmic}
\usepackage{algorithm}
\usepackage{indentfirst} 
\usepackage{cite}
\usepackage{tikz}
\usepackage{multirow}
\usepackage{mathtools}

\newcommand{\software}[1]{{\tt#1}}
\newcommand{\norm}[1]{\lVert#1\rVert}
\newcommand{\alert}[1]{{\textit{#1}}}

\begin{document}
\title{MATMPC - A MATLAB Based Toolbox for Real-time Nonlinear Model Predictive Control}
\author{Yutao Chen$^{1}$\thanks{$^{1}$Department of Information Engineering, University of Padova, Via Gradenigo, 6/B I-35131 Padova, Italy, {\tt\small \{yutao, bruschet, beghi\}@dei.unipd.it}}, Mattia Bruschetta$^{1}$, Enrico Picotti$^{1}$, Alessandro Beghi$^{1}$}

\maketitle

\begin{abstract}
In this paper we introduce \software{MATMPC}, an open source software built in \software{MATLAB} for nonlinear model predictive control (NMPC). It is designed to facilitate modelling, controller design and simulation for a wide class of NMPC applications. \software{MATMPC} has a number of algorithmic modules, including automatic differentiation, direct multiple shooting, condensing, linear quadratic program (QP) solver and globalization. It also supports a unique Curvature-like Measure of Nonlinearity (CMoN) MPC algorithm. MATMPC has been designed to provide state-of-the-art performance while making the prototyping easy, also with limited programming knowledge. This is achieved by writing each module directly in \software{MATLAB} API for C. As a result, \software{MATMPC} modules can be compiled into MEX functions with performance comparable to plain C/C++ solvers. MATMPC has been successfully used in operating systems including WINDOWS, LINUX AND OS X. Selected examples are shown to highlight the effectiveness of \software{MATMPC}.
\end{abstract}

\section{introduction}
In recent years, together with an increase of computational power, the number of applications of linear and nonlinear MPC for fast-dynamics systems has  considerably grown. While several linear MPC tools (both commercial \cite{qin2003survey, bemporad2010model} and open-source \cite{lofberg2004yalmip},) are mature and available,  the number of software for nonlinear MPC (NMPC) is rather limited \cite{qin2000overview}.


\subsection{NMPC software packages}
Existing NMPC software packages can be categorized into two main classes. The first one is characterized by software written in \software{MATLAB} and aims at algorithm development, tuning, and offline simulation, as \software{MATLAB} functions are flexible to edit and easy to understand. Popular software includes \software{GPOPS} \cite{patterson2014gpops}, \software{ICLOCS2} \cite{nie2018ICLOCS2} and \software{CasADi} \cite{andersson2018casadi}. While they are very flexible and powerful for algorithm prototyping and debugging, the computational efficiency is lost as \software{MATLAB} is not designed for computational efficiency. Therefore, it is difficult to know how efficient the NMPC algorithm is for practical applications without actually implementing it in embedded hardware.

Another class of NMPC software focus on embedded hardware and fast deployment. 
There are two main structures of such software, one based on automatic code generation and the other one employing a modular structure. The former generates a tailored piece of code of NMPC algorithm for a specific application. Software with this structure includes \software{GMRES} \cite{ohtsuka2002automatic}, \software{ACADO} \cite{houska2011auto}, \software{VIATOC} \cite{kalmari2015toolkit} and \software{Forces Pro} \cite{zanelli2017forces}. The advantage of such implementations is that the code generated is compact, self-contained and is very likely efficient and hardware compatible \cite{verschueren2018towards}. However, it lacks flexibility and maintainability since code generation is a ``black box'' to users. While code generation software is efficient and can be deployed instantly, it is not suitable for algorithm prototyping and debugging.
The latter has a modular structure, where independent algorithmic modules are implemented. Among this class, \software{ACADOS} is a C library that is computational efficient as well as flexible \cite{verschueren2018towards}. \software{CT} is a C++ library of a class of algorithms for robotic applications \cite{giftthaler2018control}. It has a number of modules including MPC and has been deployed for many real-time NMPC applications. Although efficient and useful, such software requires a decent knowledge of low-level programming languages like C/C++. In addition, additional efforts are needed to build such software at  different operating systems, in different hardware structures or high level interfaces such as \software{MATLAB} and \software{Python}.

\subsection{Features of \software{MATMPC}}
\software{MATMPC} aims at filling the gap between the two aforementioned classes of MPC software, by taking advantages from both sides. First, \software{MATMPC} is mainly written in \software{MATLAB} language and can be easily embedded in \software{SIMULINK} applications, making algorithm development and NMPC simulation extremely easy and flexible. Second, \software{MATMPC} has a modular structure and its time critical modules are written in \software{MATLAB} application program interface (API) for C. These modules are compiled into MEX functions which stand for \software{MATLAB} executable. As a result, NMPC simulation using \software{MATMPC} can achieve a competitive runtime performance against other C/C++ software. In fact, \software{MATMPC} is not a library that needs to be compiled at a given operating system before its usage, but a collection of NMPC routines that only relies on \software{MATLAB} without external library dependencies at compilation time. Each module can be replaced by another one in \software{MATMPC} or by one written by users, since modules are independent from each other.

\software{MATMPC} supports a variety of algorithms that can be easily replaced for different applications. It exploits direct multiple shooting to discretize the optimal control problem (OCP) into Nonlinear Programming problem (NLP) based on dynamic models governed by explicit or implicit ordinary differential equations. Efficient numerical integrators, e.g. explicit and implicit Runge-Kutta integrators, are implemented to approximate continuous trajectory of systems. \software{MATMPC} uses sequential quadratic programming (SQP) methods to solve the NLP. Stable condensing algorithms are employed to convert the sparse quadratic program (QP) into (partial) dense QP. A number of QP solvers are embedded, that can be selected to best fit the specific application at hand. A line search globalization algorithm is provided for searching local minimum of NLP, giving more flexibility to trade off solution accuracy and runtime performance. 

In \software{MATMPC}, a Curvature-like Measure of Nonlinearity (CMoN) SQP algorithm \cite{chen2018adaptive} is implemented. This algorithm allows to update only part of sensitivities of system dynamics between two consecutive iterations and sampling instants. The number of updated sensitivities are monitored by CMoN and automatically determined on-line, depending on how nonlinear the system is. Its control and numerical performance, including computational efficiency, robustness and convergence, is demonstrated in \cite{chen2018adaptive}.

\software{MATMPC} is open source available under \software{GPL v3} at https://github.com/chenyutao36/MATMPC.

\subsection{Paper structure}
This paper is structured as follows. Section \ref{sec2} gives an introduction of algorithms employed in \software{MATMPC}. In Section \ref{sec3}, details of modules and overall features of \software{MATMPC} are given. A nontrivial simulation example using \software{MATMPC} is described in Section \ref{sec4}, followed by the conclusions in Section \ref{sec5}.

\section{Algorithm basics}\label{sec2}
In \software{MATMPC}, a NLP is formulated by applying direct multiple shooting \cite{bock1984multiple} to an OCP over the prediction horizon $T=[t_0,t_f]$, which is divided into $N$ \emph{shooting intervals} $[t_0,t_1,\ldots,t_N]$, as follows:
\begin{subequations}\label{NLP}
\begin{align}
\min_{x_k,u_k} \,&\sum_{k=0}^{N-1} \frac{1}{2}\norm{h_k(x_k,u_k)}_W^2+\frac{1}{2}\norm{h_N(x_N)}_{W_N}^2\\
s.t.\, &0=x_0-\hat{x}_0,\label{initial value embedding}\\
&0=x_{k+1}-\phi_k(x_k,u_k),\, k=0,1,\ldots,N-1,\label{conti const}\\
&\underline{r}_k\leq r_k(x_k,u_k)\leq \overline{r}_k, \,k=0,1,\ldots,N-1,\label{path constraint}\\
&\underline{r}_N\leq r_N(x_N)\leq \overline{r}_N,
\end{align}
\end{subequations}
where $\hat{x}_0$ is the measurement of the current state. System states $x_k\in\mathbb{R}^{n_x}$ are defined at the discrete time point $t_k$ for $k=0,\ldots,N$ and the control inputs $u_k\in \mathbb{R}^{n_u}$ for $k=0,\ldots,N-1$ are piece-wise constant. Here, \eqref{path constraint} is defined by $r(x_k,u_k): \mathbb{R}^{n_x}\times\mathbb{R}^{n_u} \rightarrow \mathbb{R}^{n_r}$ and $r(x_N): \mathbb{R}^{n_x}\rightarrow \mathbb{R}^{n_l}$ with lower and upper bound $\underline{r}_k, \overline{r}_k$. Equation \eqref{conti const} refers to the \emph{continuity constraint} where $\phi_k(x_k,u_k)$ is a numerical integration operator that solves the following initial value problem (IVP) and returns the solution at $t_{k+1}$.
\begin{equation}
0=f(\dot{x}(t), x(t),u(t),t),\quad x(0)=x_k.
\end{equation}

\subsection{Sequential Quadratic Programming}
We introduce the compact notation 
\begin{equation}
    \begin{aligned}
    &\mathbf{x}= \left [x_0^\top, x_1^\top,\dots, x_N^\top\right ]^\top,\\
    &\mathbf{u}= \left [u_0^\top, u_1^\top,\dots, u_{N-1}^\top\right ]^\top
    \end{aligned}
\end{equation}
for the discrete state and control variables. Problem \eqref{NLP} is solved using SQP method, where at iteration $i$, a QP problem is formulated as

\begin{align}\label{QP}
\begin{split}
\min_{\Delta \mathbf{x},\Delta \mathbf{u}} \quad & \sum_{k=0}^{N-1}( \frac{1}{2}
\begin{bmatrix}
\Delta x_k\\
\Delta u_k
\end{bmatrix}^\top H^i_k 
\begin{bmatrix}
\Delta x_k\\
\Delta u_k
\end{bmatrix} + g_k^{i^\top}
\begin{bmatrix}
\Delta x_k\\
\Delta u_k
\end{bmatrix} ) \\
& + \frac{1}{2}\Delta x_N^\top H^i_N \Delta x_N + g_N^{i^\top}\Delta x_N \\
s.t. \quad & \Delta x_0=\hat{x}_0-x_0,\\
& \Delta x_{k+1}=A_{k}^i \Delta x_{k}+ B_{k}^i \Delta u_{k} +d_{k}^i, \\
& \underline{c}_k^i\leq C_k^i \Delta x_k + D_k^i \Delta u_k\leq \overline{c}_k^i, \\
&\underline{c}_N^i\leq C_N^i \Delta x_N \leq \overline{c}_N^i, 
\end{split}
\end{align}
where $\Delta \mathbf{x}=\mathbf{x}-\mathbf{x}^i, \Delta \mathbf{u}=\mathbf{u}-\mathbf{u}^i$ and for $k=0,1,\ldots,N-1$
\begin{align}\label{QP data}
\begin{split}
&H_k^i = \frac{\partial h_k^i}{\partial (x_k,u_k)}^\top \frac{\partial h_k^i}{\partial (x_k,u_k)}, \quad g_k^i = \frac{\partial \norm{h_k^i}_W^2}{\partial (x_k,u_k)},\\
&A_k^i=\frac{\partial \phi_k}{\partial x_k}(x_k^i,u_k^i), \quad B_k^i=\frac{\partial \phi_k}{\partial u_k}(x_k^i,u_k^i),\\
&C_k^i=\frac{\partial r_k}{\partial x_k}(x_k^i,u_k^i), \quad D_k^i=\frac{\partial r_k}{\partial u_k}(x_k^i,u_k^i),\\
&\overline{c}_k^i=\overline{r}_k-r_k(x_k^i,u_k^i),\quad  \underline{c}_k^i=\underline{r}_k-r_k(x_k^i,u_k^i),\\
&d_k^i = \phi(x_k^i,u_k^i)-x_{k+1}^i.
\end{split}
\end{align}
Here we use Gauss-Newton Hessian approximation to compute $H_k^i$ as it is a good approximate of the exact Hessian for the least square cost function in \eqref{NLP} and it is always positive semi-definite. The QP problem \eqref{QP} has a special structure and can be solved by structure exploiting or sparse solvers, such as \software{HPIPM} \cite{hpipm}, \software{OSQP} \cite{osqp}, and \software{Ipopt} \cite{wachter2006implementation}. 

An alternative is to first condense problem \eqref{QP} \cite{Andersson2013b} and obtain a dense QP problem as follows:
\begin{equation}
\begin{aligned}
\min_{\Delta \mathbf{u}} \quad& \frac{1}{2} {\Delta \mathbf{u}}^\top H_c {\Delta \mathbf{u}}+ g_c^\top \Delta \mathbf{u}\\
s.t.\quad & \underline{c}_c \leq C_c \Delta \mathbf{u} \leq \overline{c}_c,
\end{aligned}
\label{QP_full_condensed}
\end{equation}
Problem \eqref{QP_full_condensed} can be solved by dense QP solvers like \software{qpOASES} \cite{ferreau2014qpoases}. It is also possible to use partial condensing \cite{chen2018thsis} to obtain a smaller but still sparse QP problem. Computation efficiency improvement using partial condensing has been reported in \cite{frison2016efficient, chen2018efficient}.

The solution of \eqref{QP} is used to update the solution of \eqref{NLP} by
\begin{equation}
    \mathbf{x}^{i+1} = \mathbf{x}^{i} + \alpha^i \Delta\mathbf{x}^{i}, \, \mathbf{u}^{i+1} = \mathbf{u}^{i} + \alpha^i \Delta\mathbf{u}^{i},
\end{equation}
where $\alpha^i$ is the step length determined by globalization strategies. A practical line search SQP algorithm employing $\ell_1$ merit function \cite{nocedal2006numerical} is employed in \software{MATMPC}. The merit function is defined as
\begin{equation}
    m(\mathbf{w};\mu)=l(\mathbf{w})+\mu\norm{e(\mathbf{w})}_1
\end{equation}
where $\mathbf{w}=[\mathbf{x}^\top,\mathbf{u}^\top]^\top$, $l(\mathbf{w})$ is the objective function of \eqref{NLP}, $e(\mathbf{w})$ contains all constraints in \eqref{NLP} with slack variables for inequality constraints and $\mu$ the penalty parameter. The step $\alpha^i\Delta \mathbf{w}^i$ is accepted if 
\begin{equation}
    m(\mathbf{w}^i+\alpha^i\Delta \mathbf{w}^i;\mu^i)\leq m(\mathbf{w}^i;\mu^i)+\eta\alpha^i\mathcal{D}(m(\mathbf{w}^i;\mu^i);\Delta \mathbf{w}^i)
\end{equation}
where $\mathcal{D}(m(\mathbf{w}^i;\mu^i);\Delta \mathbf{w}^i)$ is the directional derivative of $m$ in the direction of $\Delta \mathbf{w}^i$. We adopt Algorithm 18.3 (\cite{nocedal2006numerical}, p. 545) to choose $\mu^i$ and compute $\alpha^i$ at each iteration. An alternative is to compute a suboptimal solution by terminating the SQP iteration early before convergence is achieved. For many applications, it is sufficient to use only one iteration with a full Newton step $\alpha=1$. Such strategy is the so-called Real-Time Iteration (RTI) scheme \cite{diehl2002real} and is supported in \software{MATMPC}.

\begin{table*}[htb]
\vspace{0.4cm}
\centering
\caption{Available options in \software{MATMPC}}
\label{option list}
\resizebox{.9\textwidth}{!}{%
\begin{tabular}{c|cccl}
\hline
Hessian Approximation       & \multicolumn{4}{c}{Gauss-Newton}                                                                                                                                                                                                                                              \\ \hline
\multirow{2}{*}{Integrator} & \multirow{2}{*}{\begin{tabular}[c]{@{}c@{}}Explicit Runge Kutta 4 \\ (\software{CasADi} code generation)\end{tabular}} & \multirow{2}{*}{Explicit Runge Kutta 4} & \multicolumn{2}{c}{\multirow{2}{*}{\begin{tabular}[c]{@{}c@{}}Implicit Runge-Kutta \\ (Gauss-Legendre)\end{tabular}}} \\
                            &                                                                                                             &                                         & \multicolumn{2}{c}{}                                                                                                  \\ \hline
Condensing                  & non                                                                                                        & full                                 & \multicolumn{2}{c}{partial}                                                                                           \\ \hline
\multirow{2}{*}{QP solver}  & \software{qpOASES}                                                                                                     & \software{MATLAB quadprog}                                & \multicolumn{2}{c}{\software{Ipopt}}                                                                                             \\
                            & \software{OSQP}                                                                                                        & \software{HPIPM}                                   & \multicolumn{2}{c}{}                                                                                                  \\ \hline
Globalization               & \multicolumn{2}{c}{$\ell_1$ merit function line search}                                                            &\multicolumn{2}{c}{Real-Time Iteration}                                                                                                                                                                                              \\ \hline
Additional features             & \multicolumn{2}{c}{CMoN-SQP}                                                                                                                          & \multicolumn{2}{c}{input MB}                                                                                          \\ \hline
\end{tabular}%
}
\end{table*}

\subsection{Curvature-like measure of nonlinearity SQP}
In \software{MATMPC}, the CMoN-SQP algorithm \cite{chen2018adaptive} is implemented to adaptively update system sensitivities on-line. The updating rule is given by
\begin{align}
\nabla \phi_{k}^{i}=\left\{
\begin{array}{l}
\nabla \phi_k^{i-1}, \: \text{if} \: \: \kappa_k^i\leq \eta^i_{pri}\, \&\, \tilde{\kappa}_k^{i}\leq \eta_{dual}^{i} , \\
\text{eval}(\nabla \phi_k^i),\, \text{otherwise}\
\end{array} \right. \label{Update Logic}
\end{align}
where $\nabla \phi_{k}=[A_k, B_k],k=0,\ldots,N-1$ is the sensitivity matrix, $(\eta^i_{pri},\eta_{dual}^{i})$ the CMoN threshold. The CMoN value is defined as
\begin{align}
&\kappa_k^{i}\coloneqq\frac{\norm{\phi_k^{i}-\phi_k^{i-1}-\nabla \phi_k^{i-1}q_k^{i-1}}}{\norm{\nabla \phi_k^{i-1}q_k^{i-1}}},\\
&\tilde{\kappa}_k^{i}\coloneqq\frac{\norm{\Delta \lambda^{i-1^\top}_{k+1}(\nabla \phi_k^{i}- \nabla \phi_k^{i-1})}}{\norm{\Delta \lambda^{i-1^\top}_{k+1} \nabla \phi_k^{i-1}}},
\end{align}
where $q_k^{i-1} = [x_k^{i}-x_k^{i-1}, u_k^{i}-u_k^{i-1}]^\top, \Delta \lambda_k^{i-1} = \lambda_k^{i}-\lambda_k^{i-1}$ are the increments on primal and dual variables between two iterations, respectively. Here, the CMoN value is an indicator of local nonlinearity of system and the updating rule \eqref{Update Logic} ensures that only sufficiently nonlinear sensitivities are updated, possibly reducing computational burden. 

CMoN-SQP only requires two user defined parameters $(\epsilon^{abs}, \epsilon^{rel})$ that are absolute and relative tolerances of the accuracy of solution of \eqref{QP}. In \cite{chen2018adaptive}, it is proved that the threshold $(\eta^i_{pri},\eta_{dual}^{i})$ is a function of $(\epsilon^{abs}, \epsilon^{rel})$. Hence, by defining the tolerance off-line, the CMoN threshold is automatically updated and the number of updated sensitivities is determined by CMoN-SQP on-line. In addition, CMoN-SQP supports the adoption of the RTI scheme by updating partial sensitivities between two sampling instants.

\section{Structure of \software{MATMPC}}\label{sec3}
\subsection{Overview}
\software{MATMPC} is a collection of \software{MATLAB} functions, including standard and MEX ones. \software{MATMPC} is an open source software (\software{GPL v3}) written in \software{MATLAB} and \software{MATLAB} C API. It consists of a number of algorithmic modules which can be easily replaced or extended. \software{MATMPC} aims at fast and flexible algorithm prototyping and competitive run-time performance.

\subsection{Modules of \software{MATMPC}}

\software{MATMPC} consists of two main functions, namely the \alert{model\_generation} and the \alert{simulation}. The \alert{model\_generation} function takes user-defined dynamic models and generates C codes of model analytic functions and their derivatives, by employing automatic differentiation (AD) in \software{CasADi} \cite{andersson2018casadi}. Note that \software{MATMPC} only generates codes from model dynamics, taking model and optimization parameters as parameters that can be altered on-line. On the other hand, other NMPC code generation tools generate ready to use codes that include the entire NMPC algorithm. Therefore, users can change model or optimization parameters on-line without repeatedly running \alert{model\_generation} function in \software{MATMPC}.

The \alert{simulation} function is for running closed-loop NMPC simulations in \software{MATLAB}. It starts from initializing controller options, data and memories defined in \software{MATLAB} \alert{struct} format. Available options in \software{MATMPC} are given in Table \ref{option list}. The NMPC controller in \software{MATMPC} is a \software{MATLAB} function that calls a number of modules. They include \alert{qp\_generation} for performing multiple shooting, \alert{condensing} for performing (partial) condensing routines, \alert{qp\_solve} for calling QP solvers, \alert{solution\_info} for computing optimal solution information such as constraint residual and Karush$-$Kuhn$-$Tucker (KKT) value, and \alert{line\_search} for performing globalization. These modules are MEX functions which share the same data and memory \alert{structs} created at initialization, without the need to allocate any memory on-line. In addition, as data memory \alert{structs} are globally accessible in \software{MATLAB}, it is possible to pause simulation and inspect intermediate data when debugging, just like running standard \software{MATLAB} functions. 

In \software{MATMPC}, there are two sources of external dependencies. The first is \software{CasADi} \cite{andersson2018casadi} for performing AD and generating C codes of model functions and derivatives. \software{CasADi} is an open source software and has pre-compiled \software{MATLAB} binaries ready to use. The second source is from QP solvers, that are carefully selected ones for NMPC applications from the open source software pool. \software{MATMPC} provides pre-compiled \software{MATLAB} binaries and interfaces for all QP solvers listed in Table \ref{option list}. While \software{qpOASES} is a dense QP solver, all the others are sparse or structure exploiting solvers. They can be called directly after multiple shooting, or after a partial condensing step that returns a smaller but also sparse QP problem. Note that these two external dependencies do not require additional compiling or installing processes.

\subsection{Features}
\software{MATMPC} has two main advantages over other NMPC software. First, the algorithm modules are written in \software{MATLAB} C API hence they can be compiled into MEX functions by using \software{MinGW} or \software{GCC}. Since these MEX functions rely on \software{MATLAB} only and not on a given operating system, \software{MATMPC} can work in WINDOWS, LINUX and OS X without any code modifications. \software{MATMPC} also employs \software{MATLAB} built-in linear algebra routines which are \software{BLAS} and \software{LAPACK} libraries from Intel MKL \cite{wang2014intel}. As a result, the compilation of MEX functions does not depend on any external header files or libraries. \software{MATMPC} is not a library but a collection of NMPC routines, each of which can be easily replaced or extended according to user needs. Second, there is no requirement on programming knowledge other than \software{MATLAB}. Users can try different combination of algorithm modules, tuning parameters and simulation modes without writing any C codes. It is also easy to replace existing modules by user defined \software{MATLAB} or MEX functions, since in \software{MATLAB}, there is no memory nor format requirement for these functions.

\subsection{Examples}
\software{MATMPC} provides several examples to illustrate its usage. The list of examples is
\begin{enumerate}
    \item Inverted Pendulum \cite{quirynen2015autogenerating}
    \item Chain of Masses (linear \cite{kirches2012efficient} and nonlinear \cite{chen2018adaptive})
    \item Hexacopter and tethered quadropter 
\end{enumerate}
These examples range from classical problems to state-of-art NMPC applications. 

\section{Co-simulation using \software{MATMPC}}\label{sec4}
We present an example of using \software{MATMPC} in a co-simulation of an automotive application: the software has been used to develop an MPC-based controller for an autonomous vehicle. The vehicle simulation model comes from a commercial simulation environment and \software{MATMPC} computes the optimal steering, throttle, and brake controls in order to follow a given trajectory.

\subsection{Control Model}
\begin{figure}
    \centering
    \includegraphics[width=\linewidth]{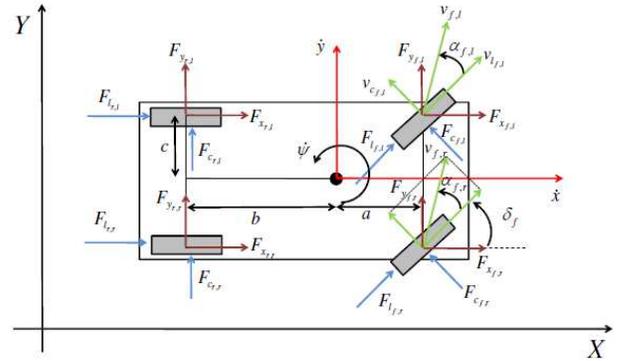}
    \caption{Forces, velocities and angles defined in the vehicle's internal model}
    \label{fig:vehicle_model}
\end{figure}

The internal model used by the controller is a four-wheel vehicle, with longitudinal forces based on a linear tire model and lateral forces based on decoupled \textit{Paceijka's magic formula} \cite{kuiper2007pac2002}. We consider the vehicle model as
\begin{equation}\label{eq:vehicle_dyn}
    \dot{\xi} = f(\xi(t), u(t)),
\end{equation}
where $\xi(t) \in \mathbb{R}^{n_x}$ is the state of the vehicle and $u(t) \in \mathbb{R}^{n_u}$ is the input. The dynamics $f$ is derived using the equation of motion of the vehicle's center of mass (CG) \cite{carvalho2013predictive}, i.e. 
\begin{equation}
\begin{aligned}
	\ddot{x} &= \dot{y}\dot{\psi} + \frac{1}{m} \left( \sum_{i,j} F_{x_{i,j}} -  F_x^d \right),\\
	\ddot{y} &= - \dot{x}\dot{\psi} + \frac{1}{m} \left( \sum_{i,j} F_{y_{i,j}} \right), \\
	\ddot{\psi} &= \frac{1}{I_z} \Bigg[ a \left( \sum_{j}F_{y_{f,j}} \right) - b \left( \sum_{j}F_{y_{r,j}} \right) \\
	&+ c \left( \sum_{i} F_{x_{i,r}}-\sum_{i}F_{x_{i,l}} \right) \Bigg]
\end{aligned}
\end{equation}
where $x,y,\psi$ are longitudinal, lateral positions and yaw angle. Subscripts $i \in \{f,r\}$ indicates front or rear wheels, $j \in \{l,r\}$ left or right wheels and $a, b, c$ are the dimensional parameters (respectively front wheels - CG longitudinal distance, rear wheels - CG longitudinal distance and wheels CG lateral distance). $F_{\{x,y\}_{\{i,j\}}}$  are the lateral and longitudinal forces on the wheels in the car reference frame and $F_x^d$ is the longitudinal drag force, detailed in Fig. \ref{fig:vehicle_model}. Finally, the \textit{slip angle} of the vehicle is defined as $\beta = atan\left(\frac{\dot{y}}{\dot{x}}\right)$.

In order to eliminate the dependency on the velocity in the reference given to the MPC, the dynamics has been reformulated in spatial coordinates w.r.t. $s$, the arc length along the track. The \textit{tracking errors} $e_{\psi} = \psi - \psi_{ref} $ and $ e_y = \norm{[X,Y]^\top-[X_{ref},Y_{ref}]^\top}_2$ are treated as additional system states. The resulting state vector is $\xi = [\dot{x}, \dot{y}, \dot{\psi}, \dot{e}_\psi, \dot{e}_y]^T$ and its derivative w.r.t $s$ is obtained using the \textit{chain rule} \cite{gao2012spatial} as 
\begin{equation}
    \xi' = \frac{d\xi}{ds} = \frac{d\xi}{dt} \frac{dt}{ds} = \frac{d\xi}{dt} \frac{1}{\dot{s}} = \frac{\dot{\xi}}{\dot{s}},
\end{equation}
where $\dot{s} = \frac{1}{1 - k \ e_y} \left(\dot{x} \ cos(e_{\psi}) - \dot{y} \ sin(e_{\psi} \right)$.
The inputs of the system are $u = [\delta_f, \gamma]^\top$, where $\delta_f$ is the steering wheel angle and $\gamma$ is the normalized throttle/braking action.

\subsection{Co-simulation environment}
\begin{figure}[!htb]
\vspace{0.4cm}
    \centering
    \includegraphics[width=\linewidth]{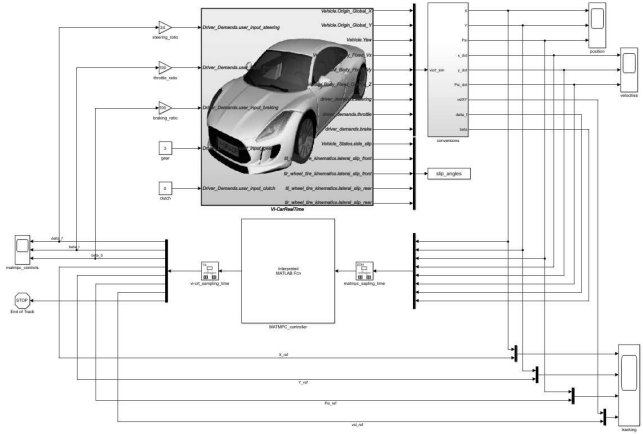}
    \caption{\software{Simulink} co-simulation block diagram}
    \label{fig:cosim}
\end{figure}

The co-simulation relied on \software{VI-CarRealTime} (\software{VI-CRT}), a simulation software specifically designed to reproduce vehicles' behaviour for high performance driving in real time \cite{vicrt}. Its simulation model has 14 degree of freedoms, 6 for the chassis and 2 for each wheel and it includes comprehensive dynamics of tire, chassis, suspensions, brakes, engine and transmission. The co-simulation is performed in \software{Simulink}, connecting a \software{VI-CRT} simulation block with \software{MATMPC} controller, as shown in Fig. \ref{fig:cosim}. In particular, \software{VI-CRT} is used to simulate at $f_s^{sim} = 1000$ Hz the dynamics of the vehicle while the control action is updated by \software{MATMPC} at $f_s^{ctrl} = 50$ Hz. The simulations have been made on a PC in WINDOWS 10, with Intel(R) Core(TM) i7-7700HQ CPU running at 2.80GHz.

\subsection{Controller setup}
The optimization problem has the form of \eqref{NLP} where the cost functions are defined as
\begin{equation}
\begin{aligned}
    & h_k(x_k,u_k) =  [ v - v_{ref}, e_y, \dot{e}_y, e_{\psi} + \beta, \dot{e}_\psi, \dot{\delta_f}, \dot{\gamma} ]^\top, \\
    & h_N(x_k,u_k) = [ v - v_{ref}, e_y, \dot{e}_y, e_{\psi} + \beta, \dot{e}_\psi ]^\top
\end{aligned}
\end{equation}
where $v=\sqrt{\dot{x}^2+\dot{y}^2}$ is vehicle's velocity. The weights are given as
\begin{equation}
\begin{aligned}
    & W = \text{diag}([ 10^0, 10^2, 10^{-2}, 10^2, 10^{-2}, 10^1, 10^{-1}]), \\
    & W_N = \text{diag}([ 10^0, 10^2, 10^{-2}, 10^2, 10^1]).
\end{aligned}
\end{equation}
The constraint functions are defined as 
\begin{equation}
\begin{aligned}
    & r_k = [e_{\psi}, e_y, \delta_f, \gamma, \dot{\delta_f}, \dot{\gamma} ]^\top, \\
    & r_N = [	e_{\psi}, e_y, \delta_f, \gamma ]^\top
\end{aligned}
\end{equation}
with bounds
\begin{equation}
\begin{aligned}    
    & \underline{r}_k = [	-\frac{\pi}{2}, -5,	-\frac{\pi}{6},	-1, -1, -5 ]^\top,\\
	& \overline{r}_k = [ +\frac{\pi}{2}, +5, +\frac{\pi}{6}, +1, +1, +5 ]^\top,\\
	& \underline{r}_N = [	-\frac{\pi}{2}, -5,	-\frac{\pi}{6},	-1 ]^\top,\\
	& \overline{r}_N = [ +\frac{\pi}{2}, +5, +\frac{\pi}{6}, +1 ]^\top.
\end{aligned}
\end{equation}

The options used in \software{MATMPC} for the simulation are summarized in Table \ref{tab:ctrl-params}. For the integrator, we employ two integration steps per shooting interval of length $T_s=2$ meters. A total number of $N=75$ shooting intervals are used, enabling a prediction length of $150$ meters on track.

\begin{table}[!ht]
\vspace{0.3cm}
\centering
\caption{\software{MATMPC} options used for the co-simulation}
\label{tab:ctrl-params}
\begin{tabular}{l|l}
\hline
              & Selected module     \\ \hline
Integrator    & Explicit Runge Kutta 4    \\
Condensing    & Non                  \\
QP Solver     & HPIPM               \\
Globalization & Real-Time Iteration \\ \hline
\end{tabular}
\end{table}

\subsection{Results using standard NMPC} \label{standard NMPC}
\begin{figure}[!ht]
\vspace{0.3cm}
    \centering
    \includegraphics[width=\linewidth]{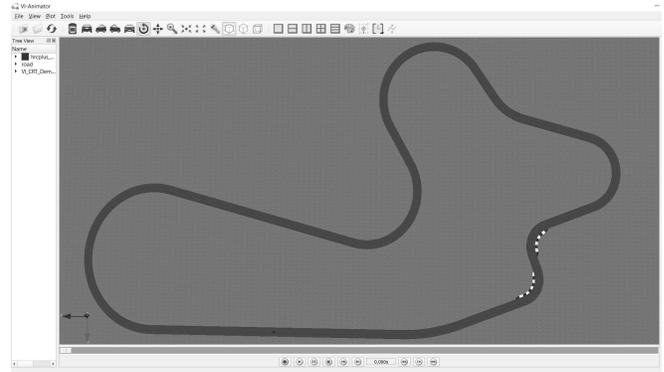}
    \caption{Simulation track in \software{VI-CRT}}
    \label{fig:sim-XY}
\end{figure}

\begin{figure}[!ht]
    \centering
    \includegraphics[width=\linewidth]{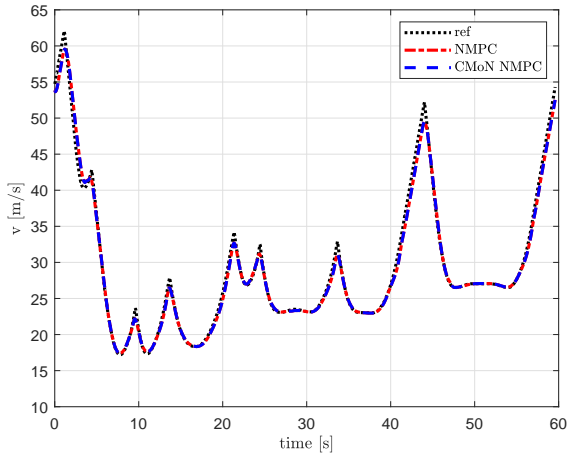}
    \caption{Simulation velocity profile and reference}
    \label{fig:sim-vel}
\end{figure}

The simulations have been performed on \textit{VI-Track} (see Fig. \ref{fig:sim-XY}), a virtual circuit available with a standard installation of \software{VI-CRT}. The reference velocity profile is obtained minimizing the lap-time by means of \software{VI-maxperf}, a tool embedded in \software{VI-CRT} that allows to compute minimum lap time simulation. The velocity profile and reference is shown in Fig. \ref{fig:sim-vel}. The MPC controller has a considerably good tracking performance while satisfying vehicles dynamics and constraints. Indeed, the \software{MATMPC} controller has been compared with the commercial controller developed by \textit{VI-Grade} that aims at driving the vehicle at the maximum performance. The \software{MATMPC} controller is able to complete the track with a smaller lap time ($T_{track}^{VI-Grade}=59.4s$ vs $T_{track}^{MPC}=59.1s$), showing superior performance of MPC on this application. The computational time for the controller is $T_{solver}^{mean} = 3.0$ ms and $T_{solver}^{max} = 10.5$ ms, showing real-time capability of \software{MATMPC} despite running in \software{MATLAB} environment.

\subsection{Results using CMoN NMPC}
\begin{figure}[!ht]
    \centering
    \includegraphics[width=\linewidth]{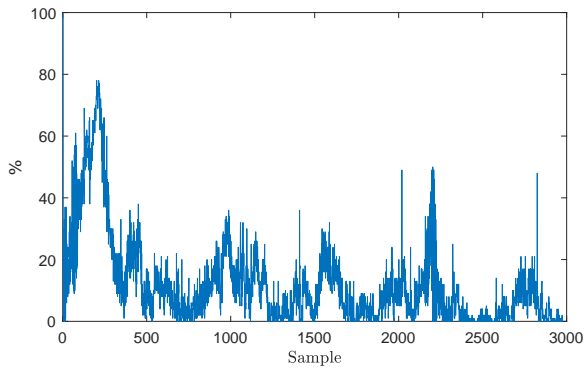}
    \caption{Percentage of exactly updated sensitivities at each sample during closed-loop simulation using CMoN-RTI scheme}
    \label{fig:perc}
\end{figure}

We also present results using the CMoN scheme, introduced in \eqref{Update Logic} from \cite{chen2018adaptive}. We use the controller configurations described in Section \ref{standard NMPC}, except for the activation of the CMoN strategy. The absolute and relative tolerance on primal and dual solutions of QP \eqref{QP} are 
$(\epsilon^{abs}, \epsilon^{rel})=(10^{-1},10^{-1})$. As can be seen in Fig. \ref{fig:sim-vel}, the tracking performance of the CMoN scheme is indistinguishable from that of the standard NMPC. However, Fig. \ref{fig:perc} shows that the percentage of exactly updated sensitivities at each sample is at most $80\%$ and in average less than $20\%$. It demonstrates the effectiveness of the CMoN scheme for a non-trivial application.

\section{Conclusion}\label{sec5}
In this paper, we introduce \software{MATMPC}, a NMPC software based on \software{MATLAB}. We present briefly the NMPC algorithm used in \software{MATMPC}, and a detailed description of the structure and features of \software{MATMPC}. Through a non-trivial vehicle control application, the effectiveness and efficiency of \software{MATMPC} is demonstrated.

\bibliographystyle{IEEEtran}
\bibliography{IEEEabrv,ecc19}
\end{document}